# Finite-time synchronization of non-autonomous chaotic systems with unknown parameters

Jianping Cai* and Meili Lin

*Department of Mathematics, Zhangzhou Normal University, Zhangzhou 363000, China*

* *Corresponding author. E-mail: mathcai@hotmail.com*

***Abstract—Adaptive control technique is adopted to synchronize two identical non-autonomous systems with unknown parameters in finite time. A virtual unknown parameter is introduced in order to avoid the unknown parameters from appearing in the controllers and parameters update laws. The Duffing equation and a gyrostat system are chosen as the numerical examples to show the validity of the present method.***

## I. INTRODUCTION

Chaos synchronization has been an interesting research area since the pioneer work by Pecora and Carroll [1], partially due to its potential applications [2, 3]. So far, most of the results on chaos synchronization focus on asymptotically stable synchronization, that is, synchronization may be achieved with infinite settling time. But in practice, one may concern firstly how to stabilize the systems within a finite-time interval. Different control methods, such as output feedback control and finite time observer, were constructed to stabilize nonlinear systems in finite time [4-10]. It was demonstrated that the finite-time control techniques have better robustness and disturbance rejection properties [11-13]. Recently finite-time control techniques have been applied in the study of chaos synchronization. An approach based upon general results borrowed from the robust control theory was used to investigate finite-time global chaos synchronization for piecewise linear maps [14]. Sliding mode control was an effective method to realize finite-time synchronization between two chaotic systems with uncertainties [15-19]. Controllers, involving the term $-ksign(x)|x|^{\alpha}$ with $0<\alpha<1$ or $e^{\beta}$ with $0<\beta=q/p<1$, were other valid methods often adopted for the studies of finite-time synchronization of two chaotic systems [20, 21]. A new method, stabilizing subsystem step by step in finite time, was proposed to study finite-time synchronization of unified chaotic system with uncertain parameters [21], and similar problem was further studied based on CLF (control Lyapunov function) and Lie derivative [22].

The uncertainties concerned in Refs.[15-22] were mostly unmatched parameters, parameter disturbances or external disturbances. So far as we know, less attention has been paid to the issue of finite-time synchronization of chaotic systems with unknown parameters. In this paper, adaptive control technique is adopted to synchronize two identical non-autonomous chaotic systems with unknown parameters in finite time. While the traditional adaptive control method is used to design proper controller to realize finite-time synchronization, we found that the controllers and parameters update laws have to contain unknown parameters, which cannot be implemented in practice. In order to overcome this difficulty, a known parameter is introduced as a virtual unknown parameter. With the help of the virtual unknown parameter, an adaptive controller and the corresponding parameters update laws are therefore designed to synchronize two coupled chaotic systems in finite time. The Duffing equation and a gyrostat system are chosen as the numerical examples to show the validity of the present method.

## II. FINITE-TIME CHAOS SYNCHRONIZATION SCHEME

Consider a class of chaotic systems with unknown parameters of the form

$$\dot{x}=f_p(t,x)+F_p(t,x)\theta_p, \tag{1}$$

where $t\in[0,\infty)$, $x\in R^n$ is the state vector, $\theta_p=(\theta_1,\theta_2,\cdots,\theta_m)^T\in R^m$ is an unknown parameter vector, $f_p(t,x)$ is a n-dimensional nonlinear function vector and $F_p(t,x)$ is a $n\times m$ nonlinear function matrix. Let $\Omega_x\subset R^n$ be a bounded region containing the whole attractor of drive system (1) such that no trajectory of system (1) ever leaves it. This assumption is simply based on the bounded property of chaotic attractor. Also, let $M\subset R^m$ be the set of parameter under which system (1) is in a chaotic state.

Introducing a virtual unknown parameter $\theta_{m+1}$, Eq.(1) can be transformed into

$$\dot{x} = f(t,x) + F(t,x)\theta \tag{2}$$

where $\theta = (\theta_1, \theta_2, \cdots, \theta_m, \theta_{m+1})^T \in R^{m+1}$ is a new parameter vector, $f(t,x)$ a n-dimensional nonlinear function vector and $F(t,x)$ is a $n \times (m+1)$ nonlinear function matrix.

Many chaotic systems of the form of system (1) can be transformed into system (2). Let us take Lorenz system as an example,

$$\begin{pmatrix} \dot{x}_1 \\ \dot{x}_2 \\ \dot{x}_3 \end{pmatrix} = \begin{pmatrix} 0 \\ -x_2 - x_1 x_3 \\ x_1 x_2 \end{pmatrix} + \begin{pmatrix} x_2 - x_1 & 0 & 0 \\ 0 & x_1 & 0 \\ 0 & 0 & x_3 \end{pmatrix} \begin{pmatrix} \theta_1 \\ \theta_2 \\ \theta_3 \end{pmatrix},$$
$$= f_p(t,x) + F_p(t,x)\theta_p$$

where $\theta_1, \theta_2, \theta_3$ are unknown parameters. By introducing a virtual unknown parameter $\theta_4 = -1$, the coefficient of $x_2$, the Lorenz system can be rewritten as

$$\begin{pmatrix} \dot{x}_1 \\ \dot{x}_2 \\ \dot{x}_3 \end{pmatrix} = \begin{pmatrix} 0 \\ -x_1 x_3 \\ x_1 x_2 \end{pmatrix} + \begin{pmatrix} x_2 - x_1 & 0 & 0 & 0 \\ 0 & x_1 & 0 & x_2 \\ 0 & 0 & x_3 & 0 \end{pmatrix} \begin{pmatrix} \theta_1 \\ \theta_2 \\ \theta_3 \\ \theta_4 \end{pmatrix}.$$
$$= f(t,x) + F(t,x)\theta$$

If system (2) is chosen as the drive system, the response system is constructed as follow,

$$\dot{y} = f(t,y) + F(t,y)\hat{\theta} + u(t), \tag{3}$$

where $y \in R^n$ is the state vector, $f(t,y)$ is a n-dimensional nonlinear function vector and $F(t,y)$ is a $n \times (m+1)$ nonlinear function matrix., $\hat{\theta} \in R^{m+1}$ is a adaptive parameter vector and $u(t)$ is a controller. Let $\Omega_y \subset R^n$ be a bounded region containing the whole attractor of response system (3) without control.

The object of this paper is to design a proper adaptive controller $u(t)$ such that the response system (3) will synchronize the drive system (2) in finite time. To this end, some hypotheses are needed:

**H1.** The nonlinear functions $f(t,\cdot)$ and $F(t,\cdot)$ are continuous on a bounded closed region $[0,\infty) \times \Omega$ and $\Omega$ containing both $\Omega_x$ and $\Omega_y$.

**H2.** The nonlinear matrix function $f(t,\cdot)$ satisfies Lipschitz condition in the region $[0,\infty) \times \Omega$,

$$\|f(t,x) - f(t,y)\| \le L\|x - y\|,$$

where $L$ is an appropriate positive constant. In this paper, $\|\cdot\|$ denotes matrix or vector norm, defined as $\|A\| = (\sum_{j=1}^{m}\sum_{i=1}^{n} a_{ij}^2)^{1/2}$ for the matrix $A = (a_{ij})_{n \times m}$.

**H3.** The unknown parameter $\theta = (\theta_1, \theta_2, \cdots, \theta_m, \theta_{m+1})^T$ is norm bounded. Suppose that

$$|\theta_i| \le d_i,\ i = 1, 2, \cdots, m+1,$$

then

$$\|\theta\| \le d,$$

where $d = (d_1^2 + d_2^2 + \cdots + d_{m+1}^2)^{1/2}$ are positive constants.

Define the error variable as $e = x - y$. Subtracting (3) from (2) yields the synchronization error dynamics as

$$\dot{e} = f(t,x) - f(t,y) + F(t,x)\theta - F(t,y)\hat{\theta} - u(t) \tag{4}$$

Now the finite-time synchronization between systems (2) and (3) is transformed into the finite-time stability of error dynamics (4), which means that there exists a constant $T > 0$ such that

$$\lim_{t \to T}\|e\| = \lim_{t \to T}\|x - y\| = 0,$$

and $\|e\| = \|x - y\| \equiv 0$, for $t \ge T$. Some necessary lemmas are introduced in the follows.

**Lemma 1** [21] Assume that a continuous, positive-definite function $V(t)$ satisfies the following differential inequality,

$$\dot{V}(t) \le -cV^{\eta}(t),\ \text{for any } t > 0, V(t_0) \ge 0,$$

where $c > 0, 0 < \eta < 1$ are all constants. Then, for any given $t_0$, $V(t)$ satisfies the following inequality,

$$V^{1-\eta}(t) \le V^{1-\eta}(t_0) - c(1-\eta)(t - t_0), t_0 \le t \le t_1,$$

and $V(t) \equiv 0$ for $t > t_1$ with $t_1 = t_0 + \dfrac{V^{1-\eta}(t_0)}{c(1-\eta)}$.

Based on Lemma 1 and motivated by the Theorem 4.10 in [24], we establish a lemma for finite-time stable of non-autonomous systems.

Consider the non-autonomous system

$$\dot{z} = g(t,z) \tag{5}$$

where $g: [0,\infty) \times D \to R^n$ is piecewise continuous in $t$ and locally Lipchitz in $z$ on $[0,\infty) \times D$, and $D \subset R^n$ is a domain that contains the origin $z = 0$. The origin is an equilibrium point for Eq.(5) at $t = 0$ if $g(t,0) = 0$, for $t \ge 0$.

**Lemma 2** Let $z = 0$ be an equilibrium point for Eq.(5) and $D \subset R^n$ be a domain containing the origin $z = 0$. Let $V: [0,\infty) \times D \to R^n$ be a continuously differentiable function such that

$$a_1\|z\|^p \le V(t,z) \le a_2\|z\|^p, \tag{6}$$

$$\frac{\partial V}{\partial t} + \frac{\partial V}{\partial z}g(t,z) \le -a_3\|z\|^q, \tag{7}$$

for any $t \ge 0$ and $z \in D$, where $a_1, a_2, a_3, p$ and $q$ are positive constants and $\dfrac{q}{p} < 1$. Then, $z = 0$ is finite-time stable. If the assumptions hold globally, then $z = 0$ is globally finite-time stable.

**Proof:** The derivative of $V$ along the trajectories of (5) is given by

$$\dot{V}(t,z)=\frac{\partial V}{\partial t}+\frac{\partial V}{\partial z}g(t,z)\le -a_3\|z\|^q\le 0 .$$

By the inequalities (6), trajectories starting in $\{a_2\|z\|^p\le h\}\subset\{V(t,z)\le h\}$, for sufficiently small $h$, remain bounded for all $t\ge t_0$. Inequalities (6) and (7) show that $V$ satisfies the differential inequality

$$\dot{V}\le -a_3\|z\|^q=-a_3(\|z\|^p)^{\frac{q}{p}}\le -a_3(\frac{V}{a_2})^{\frac{q}{p}}=-a_3a_2^{-\frac{q}{p}}V^{\frac{q}{p}} ,$$

$$V^{1-\frac{q}{p}}(t)\le V^{1-\frac{q}{p}}(t_0)-a_3a_2^{-\frac{q}{p}}(1-\frac{q}{p})(t-t_0),\quad t_0\le t\le t_1 ,$$

and $V(t)\equiv 0$, for $t\ge t_1$ with $t_1=t_0+\dfrac{a_2^{\frac{q}{p}}V^{1-\frac{q}{p}}(t_0)}{a_3(1-\frac{q}{p})}$.

By the inequalities (6),

$$\|z\|\le(\frac{V(t,z)}{a_1})^{\frac{1}{p}} ,$$

so $z(t)\equiv 0$, for $t\ge t_1$. Thus, the origin is finite-time stable. If the assumptions hold globally, $h$ can be chosen arbitrarily large and the foregoing inequality holds for all $z(t_0)\in R^n$.

**Theorem 1** The coupled system (2) and (3) can be synchronized in finite time if the hypotheses H1~H3 hold and the following conditions (I)-(II) are satisfied,

(I) The controller $u(t)$ is chosen as

$$u(t)=k_1e+\frac{k_2e}{\|e\|}+F(t,x)\hat{\theta}-F(t,y)\hat{\theta}+\frac{2k_2(d^2+d\|\hat{\theta}\|)}{|\theta_{m+1}-\hat{\theta}_{m+1}|}\cdot\frac{e}{\|e\|^2} \tag{8}$$

where $k_1\ge L$, $k_2>0$.

(II) The update law of the parameter $\hat{\theta}$ is

$$\dot{\hat{\theta}}=F(t,x)^Te+\frac{k_2(\Delta-\hat{\theta})}{|\theta_{m+1}-\hat{\theta}_{m+1}|} , \tag{9}$$

where $\Delta=(d_1,d_2,\cdots,d_{m+1})^T$ and $\hat{\theta}_{m+1}$ is the $(m+1)$-th component of vectors $\hat{\theta}$.

For a fixed $t_0$, the finite time $t_1$ is determined by

$$t_1=t_0+\frac{2}{k_2}V^{\frac{1}{2}}(t_0) . \tag{10}$$

Where $V$ is a positive-definite function satisfying Lemma 2.

**Proof**: Choose a Lyapunov function of the form

$$V(e,\tilde{\theta})=\frac{1}{2}e^Te+\frac{1}{2}\tilde{\theta}^T\tilde{\theta} ,$$

where $\tilde{\theta}=\theta-\hat{\theta}$. In fact, $V(e,\tilde{\theta})=\frac{1}{2}(\|e\|^2+\|\tilde{\theta}\|^2)$ satisfies Eq.(6) in Lemma 2 in the case of $z=(e,\tilde{\theta})$, that is,

$$\frac{1}{4}(\|e\|^2+\|\tilde{\theta}\|^2)\le V(e,\tilde{\theta})\le\|e\|^2+\|\tilde{\theta}\|^2 . \tag{11}$$

The time-derivative of $V$ along the error system (4) is

$$\begin{aligned}\dot{V}&=e^T\dot{e}-(\theta-\hat{\theta})^T\dot{\hat{\theta}}\\&=e^T(f(t,x)-f(t,y)+F(t,x)\theta-\\&\quad F(t,y)\hat{\theta}-u(t))-(\theta-\hat{\theta})^T\dot{\hat{\theta}}\end{aligned}$$

From above hypotheses H1~H3, Eqs.(8) and (9), one can obtain

$$\begin{aligned}\dot{V}&=e^T(f(t,x)-f(t,y)+F(t,x)\theta-F(t,y)\hat{\theta}\\&\quad -k_1e-\frac{k_2e}{\|e\|}-F(t,x)\hat{\theta}+F(t,y)\hat{\theta}-\\&\quad \frac{2k_2(d^2+d\|\hat{\theta}\|)}{|\theta_{m+1}-\hat{\theta}_{m+1}|}\cdot\frac{e}{\|e\|^2})-(\theta-\hat{\theta})^T(F(t,x)^Te\\&\quad +\frac{k_2(\Delta-\hat{\theta})}{|\theta_{m+1}-\hat{\theta}_{m+1}|})\\&\le L\|e\|^2-k_1\|e\|^2-k_2\|e\|\\&\quad -\frac{2k_2(d^2+d\|\hat{\theta}\|)}{|\theta_{m+1}-\hat{\theta}_{m+1}|}-\frac{k_2(\theta-\hat{\theta})^T(\Delta-\hat{\theta})}{|\theta_{m+1}-\hat{\theta}_{m+1}|}\end{aligned}$$

By using $k_1\ge L$, it yields that

$$\begin{aligned}\dot{V}&\le(L-k_1)\|e\|^2-k_2\|e\|-\frac{k_2(\theta-\hat{\theta})^T(\theta-\hat{\theta})}{|\theta_{m+1}-\hat{\theta}_{m+1}|}\\&\quad +\frac{k_2(\theta-\hat{\theta})^T(\theta-\Delta)}{|\theta_{m+1}-\hat{\theta}_{m+1}|}-\frac{2k_2(d^2+d\|\hat{\theta}\|)}{|\theta_{m+1}-\hat{\theta}_{m+1}|}\end{aligned} \tag{12}$$

Using the inequalities

$$(\theta-\hat{\theta})^T(\theta-\Delta)\le 2(d^2+d\|\hat{\theta}\|)$$

and

$$|\theta_{m+1}-\hat{\theta}_{m+1}|\le\|\theta-\hat{\theta}\| ,$$

we have

$$\dot{V}\le -k_2(\|e\|+\|\theta-\hat{\theta}\|)=-k_2(\|e\|+\|\tilde{\theta}\|)\le -k_2\sqrt{\|e\|^2+\|\tilde{\theta}\|^2} \tag{13}$$

The equation (13) shows that the inequality (7) in Lemma 2 is fit. By Lemma 2 the error system (4) is stabilized at origin point $e=0$ in finite time. Furthermore the settling time $t_1$ can be defined by

$$t_1=t_0+\frac{V^{1-\frac{1}{2}}(t_0)}{k_2(1-\frac{1}{2})}=t_0+\frac{2}{k_2}V^{\frac{1}{2}}(t_0) .$$

That is, the response system (3) synchronizes the drive system (2) in finite time.

**Remark 2.** If we use the same method in [23] instead of the virtual unknown parameter, the controller and parameter update laws should be

$$u(t) = k_1 e + \frac{k_2 e}{\|e\|} + F(x)\hat{\theta} - F(y)\hat{\theta} + \frac{2k_2(d^2 + d\|\hat{\theta}\|)}{|\theta_i - \hat{\theta}_i|} \cdot \frac{e}{\|e\|^2},$$

$$\dot{\hat{\theta}} = F(x)^T e + \frac{k_2(\Delta - \hat{\theta})}{|\theta_i - \hat{\theta}_i|},$$

which contain unknown parameters $\theta_i$ $(1 \le i \le m)$ of the drive system. Theorem 1 indicates that introduction of a virtual unknown parameter can avoid the unknown parameters from appearing in controllers and parameters update laws and realize the finite-time synchronization.

**Remark 3.** The magnitude of $e/\|e\|^2$ in the controller $u(t)$ will turn to infinity as $e \to 0$. So, we take $e/(\|e\|^2 + \varepsilon)$ instead of $e/\|e\|^2$ in practice with $\varepsilon$ a sufficient small positive constant. Such substitution is also used in [25].

## III. NUMERICAL SIMULATION

The Duffing equation and a gyrostat system are selected as examples to verify the validity of the proposed method, respectively.

**Example 1** The classical Duffing equation with some unknown parameters is assumed to be

$$\ddot{x} + \theta_1 \dot{x} - x + x^3 = \theta_2 \cos \omega t, \tag{14}$$

where $\theta_1, \theta_2$ and $\omega$ are unknown parameters. Under some parameter values $\theta_1 = 0.25$, $\theta_2 = 0.4$ and $\omega = 1$, the Duffing equation exhibits chaos behavior. Let $x_1 = x, x_2 = \dot{x}, x_3 = \omega t$, and rewrite Eq.(14) in vector form

$$\begin{pmatrix} \dot{x}_1 \\ \dot{x}_2 \\ \dot{x}_3 \end{pmatrix} = \begin{pmatrix} x_2 \\ x_1 - x_1^3 \\ 0 \end{pmatrix} + \begin{pmatrix} 0 & 0 & 0 \\ -x_2 & \cos x_3 & 0 \\ 0 & 0 & 1 \end{pmatrix} \begin{pmatrix} \theta_1 \\ \theta_2 \\ \omega \end{pmatrix}.$$

By introducing a virtual unknown parameter $\theta_3 = -1$, the coefficient of $x_1^3$, the Duffing equation can be rewritten as

$$\begin{pmatrix} \dot{x}_1 \\ \dot{x}_2 \\ \dot{x}_3 \end{pmatrix} = \begin{pmatrix} x_2 \\ x_1 \\ 0 \end{pmatrix} + \begin{pmatrix} 0 & 0 & 0 & 0 \\ -x_2 & \cos x_3 & x_1^3 & 0 \\ 0 & 0 & 0 & 1 \end{pmatrix} \begin{pmatrix} \theta_1 \\ \theta_2 \\ \theta_3 \\ \omega \end{pmatrix}. \tag{15}$$

We chose Eq.(15) as the drive system and the response system is constructed as

$$\begin{pmatrix} \dot{y}_1 \\ \dot{y}_2 \\ \dot{y}_3 \end{pmatrix} = \begin{pmatrix} y_2 \\ y_1 \\ 0 \end{pmatrix} + \begin{pmatrix} 0 & 0 & 0 & 0 \\ -y_2 & \cos y_3 & y_1^3 & 0 \\ 0 & 0 & 0 & 1 \end{pmatrix} \begin{pmatrix} \hat{\theta}_1 \\ \hat{\theta}_2 \\ \hat{\theta}_3 \\ \hat{\omega} \end{pmatrix} + \begin{pmatrix} u_1 \\ u_2 \\ u_3 \end{pmatrix}. \tag{16}$$

The controllers $u_1(t), u_2(t)$ and $u_3(t)$ are defined by Eq.(8). In simulation, values of uncertain parameters of the drive system are chosen as $\theta_1 = 0.25, \theta_2 = 0.4$, $\theta_3 = -1$ and $\omega = 1$. The bounds of unknown parameters are assumed to be $d_1 = 0.5, d_2 = 1, d_3 = 1$ and $d_4 = 2$. The small positive constant is set to be $\varepsilon = 0.0001$, and $k_1 = 1$ and $k_2 = 0.0001$. The initial values are $(x_1(0), x_2(0), x_3(0)) = (1,1,1)$, $(y_1(0), y_2(0), y_3(0)) = (-2,3,-1)$, $(\hat{\theta}_1(0), \hat{\theta}_2(0), \hat{\theta}_3(0), \hat{\omega}(0)) = (0.1, 0.2, 0.5, 0.3)$, respectively. From Fig.1, one can see the error between the drive and response systems $\|e\| = \sqrt{x_1^2 + x_2^2 + x_3^2}$ converges to zero at about $t = 2$. The time histories of controllers $u_1(t), u_2(t)$ and $u_3(t)$ are shown in Fig.2.

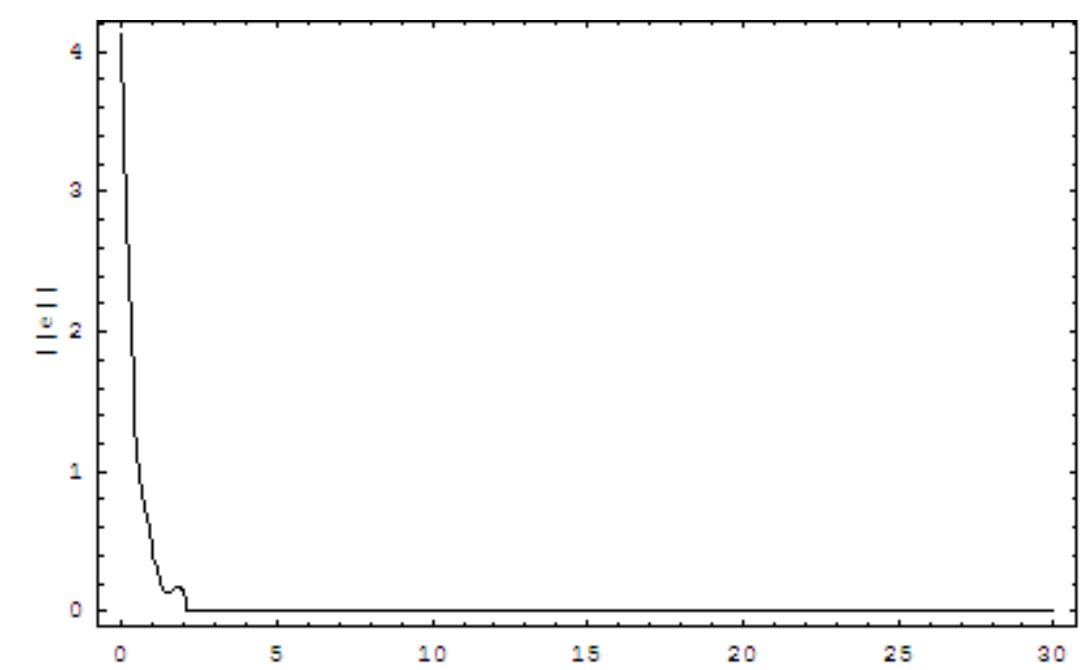

Fig.1 Error between the drive system (15) and response system (16)

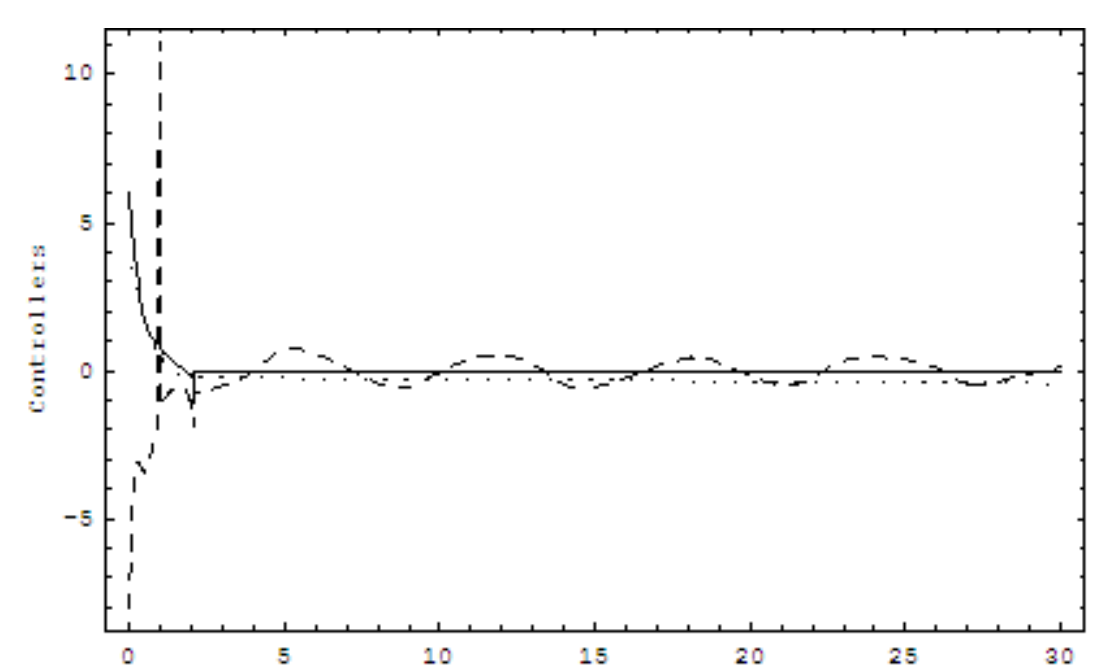

Fig.2 Time histories of controllers in Eq.(16), solid curve for $u_1(t)$, dashed curve for $u_2(t)$ and dotted curve for $u_3(t)$

**Example 2** A gyrostat system with some unknown parameters is assumed to be

$$\dot{x}_1 = -x_2x_3 - (a_1 + a_2\cos t)x_2 + 0.4x_3 - 0.002(x_1 + x_1^3)$$
$$\dot{x}_2 = x_1x_3 - 0.4x_3 + (a_1 + a_2\cos t)x_1 - 0.002(x_2 + x_2^3) \qquad (17)$$
$$\dot{x}_3 = -0.2x_1 + 0.2x_2 - 0.2x_3 - 0.001(x_3 + x_3^2) + a_3\sin t$$

where $a_1, a_2$ and $a_3$ are unknown parameters. Under some parameter values, the gyrostat system exhibits chaos behavior. Readers refer to Ref.[26] for more details. Introduce a virtual unknown parameter $a_4 = 0.4$ , the coefficient of $x_3$ in the first equation in Eq.(17), and rewrite it in vector form

$$\begin{pmatrix}\dot{x}_1\\ \dot{x}_2\\ \dot{x}_3\end{pmatrix} = \begin{pmatrix}-x_2x_3 - 0.002(x_1 + x_1^3)\\ x_1x_3 - 0.4x_3 - 0.002(x_2 + x_2^3)\\ -0.2x_1 + 0.2x_2 - 0.2x_3 - 0.001(x_3 + x_3^2)\end{pmatrix} + \begin{pmatrix}-x_2 & -x_2\cos t & 0 & x_3\\ x_1 & x_1\cos x_3 & 0 & 0\\ 0 & 0 & \sin t & 0\end{pmatrix}\begin{pmatrix}a_1\\ a_2\\ a_3\\ a_4\end{pmatrix} \qquad (18)$$

We chose Eq.(17) as the drive system and the response system is constructed as

$$\begin{pmatrix}\dot{y}_1\\ \dot{y}_2\\ \dot{y}_3\end{pmatrix} = \begin{pmatrix}-y_2y_3 - 0.002(y_1 + y_1^3)\\ y_1y_3 - 0.4y_3 - 0.002(y_2 + y_2^3)\\ -0.2y_1 + 0.2y_2 - 0.2y_3 - 0.001(y_3 + y_3^2)\end{pmatrix} + \begin{pmatrix}-y_2 & -y_2\cos t & 0 & y_3\\ y_1 & y_1\cos x_3 & 0 & 0\\ 0 & 0 & \sin t & 0\end{pmatrix}\begin{pmatrix}\hat{a}_1\\ \hat{a}_2\\ \hat{a}_3\\ \hat{a}_4\end{pmatrix} + \begin{pmatrix}u_1(t)\\ u_2(t)\\ u_3(t)\end{pmatrix} \qquad (19)$$

In simulation, values of uncertain parameters of the drive system are chosen as $a_1 = 0.5, a_2 = 3.25$ and $a_3 = 1.625$ . The bounds of unknown parameters are assumed to be $d_1 = 1, d_2 = 4, d_3 = 2$ and $d_4 = 0.4$ . The small positive constant is set to be $\varepsilon = 0.0001$ , and $k_1 = 4$ and $k_2 = 0.0001$ . The initial values are $(x_1(0), x_2(0), x_3(0)) = (1,1,1)$ , $(y_1(0), y_2(0), y_3(0)) = (-1,-1,-1)$ , $(\hat{a}_1(0), \hat{a}_2(0),$ $(\hat{a}_1(0), \hat{a}_2(0), \hat{a}_3(0), \hat{a}_4(0)) = (0.1, 2, 0.5, 0.1)$ , respectively. From Fig.3, one can see the error between the drive and response systems $\|e\|$ converges to zero at about $t = 2.2$ . The time histories of controllers $u_1(t)$, $u_2(t)$ and $u_3(t)$ are shown in Fig.4.

## IV. CONCLUSION

This work presents a general method for synchronizing two identical non-autonomous chaotic systems with unknown parameters in finite time. The introduction of a virtual unknown parameter can avoid the unknown parameters from appearing in controllers and parameters update laws. Numerical simulations on the basis of the Duffing equation and a gyrostat system show that the proposed method is effective.

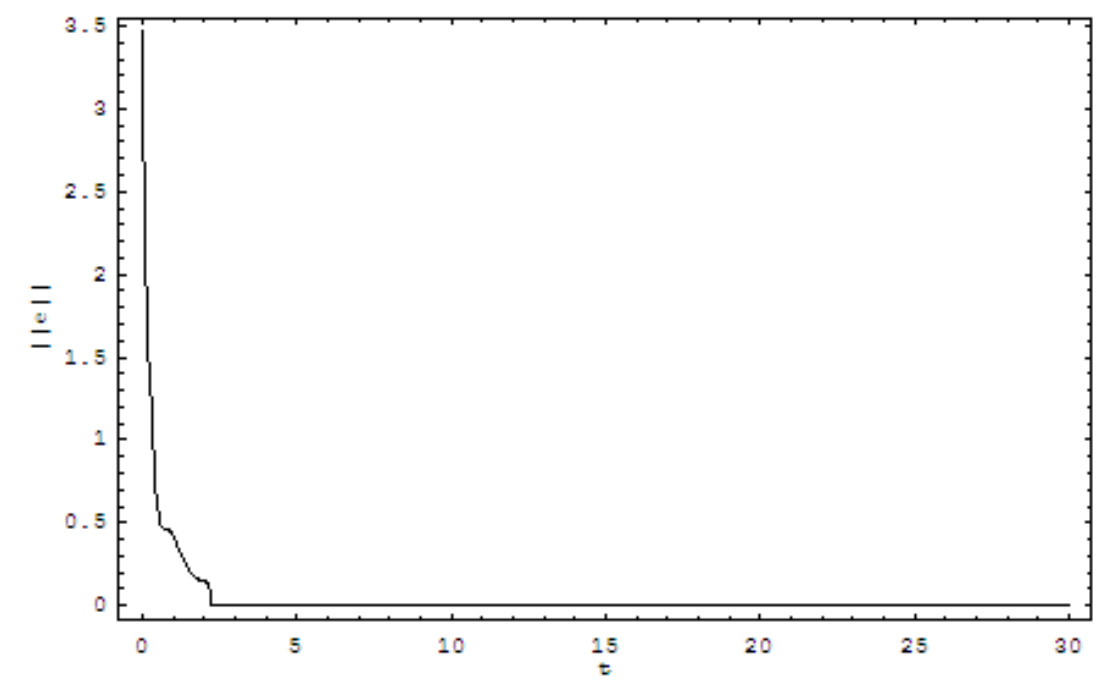

Fig.3 Error between the drive system (18) and response system (19)

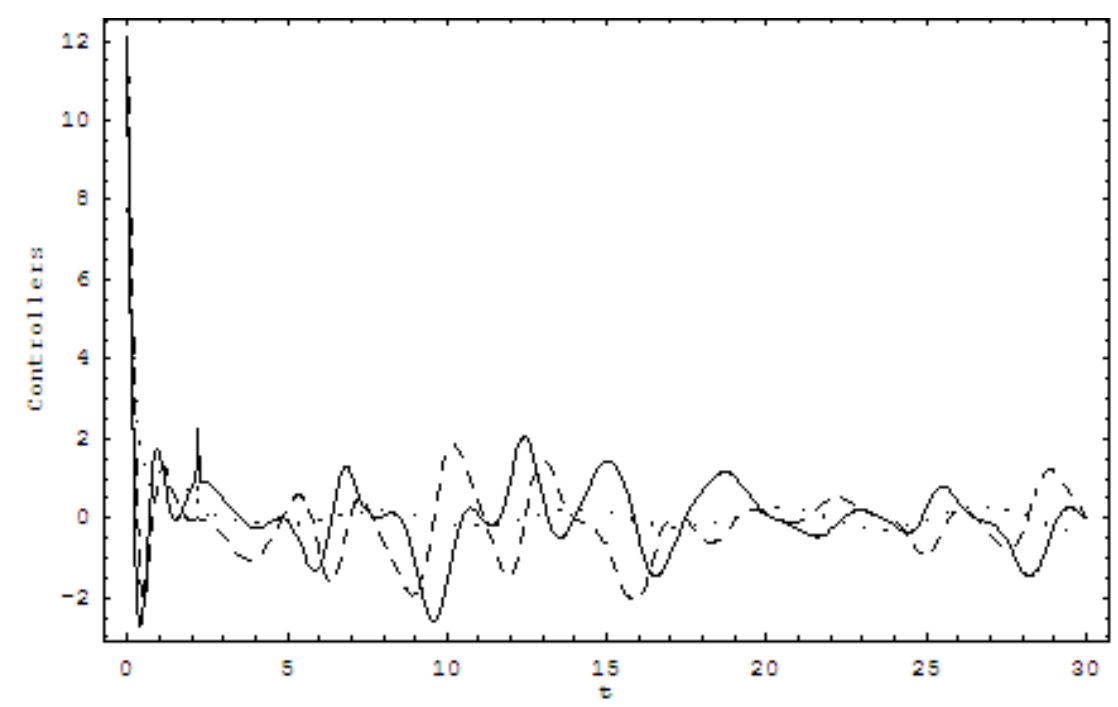

Fig.4 Time histories of controllers in Eq.(19), solid curve for $u_1(t)$ , dashed curve for $u_2(t)$ and dotted curve for $u_3(t)$

## ACKNOWLEDGMENT

Research is supported by the Foundation for Supporting Universities in Fujian Province of China under grant No JK2009020.